\documentclass[12pt,preprint]{aastex}
\def\fz{photometric redshifts }
\def\fzz{photometric redshifts}

\def\k{K_{AB}}
\def\gtsima{$\; \buildrel > \over \sim \;$}
\def\gsim{\lower.5ex\hbox{\gtsima}}

\lefthead{Fontana et al.}
\righthead{The assembly of massive galaxies in the HDFS}

\begin{document}
   \title{The assembly of massive galaxies from NIR observations of the Hubble Deep Field South.}

\author{A. Fontana$^1$, I. Donnarumma$^1$, E. Vanzella$^2,3$,
E. Giallongo$^1$, N. Menci$^1$, M. Nonino$^4$, P. Saracco$^5$,
S. Cristiani$^4$, S. D'Odorico$^2$, F. Poli$^1$}

\begin{abstract}

We use a deep $K_{AB}\leq 25$ galaxy sample in the Hubble Deep Field
South to trace the evolution of the cosmological stellar mass density
from $z\simeq 0.5$ to $z\simeq 3$.  We find clear evidence for a
decrease of the average stellar mass density at high redshift, $2\leq
z \leq 3.2$, that is $15^{+25}_{-5}\%$ of the local value, two times
higher than what observed in the Hubble Deep Field North.  To take
into account for the selection effects, we define a homogeneous
subsample of galaxies with $10^{10}M_\odot \leq M_* \leq
10^{11}M_\odot$: in this sample, the mass density at $z>2$ is
$20^{+20}_{-5} \%$ of the local value.  In the mass--limited subsample
at $z>2$, the fraction of passively fading galaxies is at most 25\%,
although they can contribute up to about 40\% of the stellar mass
density. On the other hand, star--forming galaxies at $z>2$ form stars
with an average specific rate at least $<\dot M/M_*> \simeq 4\times 10^{-10}
$~yr$^{-1}$, 3 times higher than the $z\leq 1$ value. This
implies that UV bright star--forming galaxies are substancial contributors 
to the rise of the stellar mass density with cosmic time.
Although these results are globally consistent with $\Lambda$--CDM
scenarios, the present rendition of semi analytic models fails to
match the stellar mass density produced by more massive galaxies
present at $z>2$.

\end{abstract}


\altaffiltext{1}{INAF, Osservatorio Astronomico di Roma, via Frascati 33,
I-00040, Monteporzio, Italy} 

\altaffiltext{2}{European Southern Observatory, Karl-Schwarzschild-Strasse 2,
D-85748, Garching, Germany} 

\altaffiltext{3}{INAF, Osservatorio Astronomico di Padova, Vicolo dell'Osservatorio 2, I-35122 Padova, Italy}

\altaffiltext{4}{INAF, Osservatorio Astronomico di Trieste, via
G.B. Tiepolo 11, I-34131, Trieste, Italy}

\altaffiltext{5}{INAF, Osservatorio Astronomico di Brera, via
E. Bianchi, 46 Merate, Italy}

%

\section{Introduction}

The most distintive features of current theories of galaxy formation
is the mechanism relating the assembly of the galactic potential wells
(mainly determined by the dark matter content) with the build-up of
the stellar population contained therein.  Hierarchical
theories of galaxy formation are characterized by a gradual enrichment
of the star content of galaxies driven by their progressive growth
through merging events. In turn, this implies an appreciable antievolution
with $z$ of the galacytic stellar mass distribution, in particular for
the massive galaxies which, in hierarchical theories, have assembled
relatively late.

Recent surveys have started to directly follow this process by
estimating the stellar mass content of galaxies up to $z\sim 3$,
either from detailed spectral analysis (Kauffmann et al 2002) or from
multiwavelenght imaging observations (Giallongo et al 1998, Brinchmann
and Ellis 2000, BE00 hereafter, Drory et al 2001, Cole et al 2001,
Papovich et al 2001, P01 hereafter, Shapley et al 2001, Dickinson et
al 2003, D03 hereafter). The latter technique relies on multicolor
broad band imaging - extended into the near--IR range - to estimate
the stellar content by a comparison with spectral synthesis models.
In this work we present the results that this technique yields when
applied to the Hubble Deep Field South (HDFS) data at $0<z<3.2$, and
compare them with the predictions of a semi-analytic model in a
$\Lambda$-dominated cosmology ($\Omega_m =0.3$, $\Omega_\Lambda =0.7$
and $H_0 = 70$ km s$^{-1}$Mpc$^{-1}$).  A Salpter IMF is used in the
paper.

\section{Stellar masses from the HDFS data}

The data that we use here are the results of the HDFS multiwavelength
deep imaging survey, that combines the HST optical data
(Casertano et al 2000) in the F300W, F450W, F606W and F814W filters
(hereafter $U$,$B$,$V$ and $I$) and the ultradeep VLT-ISAAC IR images
in the $Js$, $H$ and $Ks$ filters (the AB photometric system has been
used throughout the paper). The latter data, that are in common with
the FIRES survey (Labbe' et al 2002, Franx et al. 2003), have typical
exposure times of about 30 hr in each filter and were reduced and
analyzed following the recipes described in Vanzella et al 2001 and
Fontana et al 2000(F00 hereafter). An a posteriori correlation with
the FIRES data shows a good agreement in the measured colors and noise
estimates.  We will use here the $\k\leq25$ sample, where the $S/N$
ratio is larger than 5, although our results rely essentially on the
brighter objects.  The sample contains $302$ objects, $59$ of which
with spectroscopic redshift (Vanzella et al. 2002 and Sawicki et al
2003), that were used to verify that the accuracy of \fz is comparable
to our results in the HDFN.

On this catalog, we have estimated the stellar content of the HDFS
galaxies by a comparison with spectral synthesis models, following the
recepies of previous works (Giallongo et al. 1998, BE00, F00, P01, D03), In particular, we have adopted the same
parametrization: we have used the Bruzual \& Charlot (1993)
GISSEL 2000 code with Salpter IMF, a range of exponentially declining
star--formation histories with timescales from $\tau=0.1$ Gyrs to
$\tau=\infty$ Gyrs, metallicities of $Z/Z_\odot=0.02,0.2,1,2.5$, and
dust extinction ($0\leq E(B-V) \leq 1$). As in F00, we have also added
a set of multiple burst models, although only 5\% of the objects
turned out to be fitted by such models.  The
best--fitting model to the observed multiwavelength distribution (at
the spectroscopic or \fzz) is used to estimate the stellar mass $M_*$
(which includes a correction for the recycled fraction) in each galaxy
sample. As in P01, we tested both a SMC-like extinction curve (used as
attenuation) and a Calzetti 2000 one. With the latter, the
mass estimates are slightly smaller than SMC (25\% on average),
since the Calzetti curve yields lower fitted ages than the SMC, with
an unpleasant fraction of 40\% of galaxies wit ages $\leq$ 0.1 Gyrs at
any $z$. The fits based on the Calzetti curve  have also typically
poorer $\chi^2$, and would be preferred only for 20\%(40\%) of the
$z\leq2$( $z>2$) galaxies.  Given the uncertainities in the dust
treatment, and due to these systematic differences, we present both
results separately, with more emphasis on the SMC--based results.

The uncertainties involved in this approach, due to the degeneracy
among the input models as well as from photometric noise and from the
use of \fzz, have been estimated on the basis of the reduced
chi--square $\chi^2$, computed as in F00.  The $1\sigma$ confidence
levels on the fitted parameters (such as mass, age and star--formation
rate) have been obtained by scanning the model grid and retaining only
the models that have $\chi^2\leq\chi^2_{bestfit}+1$.  Prior to this,
as in P01, we have rescaled the noise in bright objects order to have
$\chi^2_{bestfit}=1$ The scan is performed either at fixed redshift
(for objects with known spectroscopic redshift) or allowing the models
to move around the best--fitting \fzz.

In the following, we shall also make use of a ``Maximal Mass''
estimate, to provide upper limits to the estimated stellar mass of
each object, being still broadly consistent with the observed
colors. We have first assumed that all the UV light is due to a recent
starburst with minimal $M_*/L$ ratio (obtained from a model with
constant SFR, $Z=0.2Z_\odot$, age$=0.2$Gyrs) and computed its
contribution to the $K$ band flux. Then, we have used a maximally old
stellar population to convert the residual $K$ band to $M_*$.

Finally we remind that, {\it IR--selected samples do not strictly
correspond to mass--selected ones}.  At any Hubble time
(i.e. redshift), the faint side of the sample is biased against the
detection of high--mass, low--luminosity objects, i.e.  old, passively
evolving or highly extincted galaxies.  Because of the uncertainties
in the modelling of such objects, it is difficult to define a clear
selection threshold.  We plot in Fig.1 two selection curves, such that
objects above the lines should be used to compute statistics that
require mass--selected samples, although several objects of lower mass
(i.e. star--forming galaxies with lower $M_*/L$) are detected.  In one
case we use a GISSEL 2000 dust-free passively evolving model ignited
by a histantaneous burst of sub--solar metallicity at $z=18$
(short-dashed line); in another we use the more realistic model of Le
Borgne and Rocca--Volmerange 2002 (long-dashed line) that reproduces
the colors of local E0 and that self--consistently includes the
effects of finite burst duration, dust--absorption and metallicity
evolution. 

Our sample is definitely incomplete below these curves, and reasonably
complete above, except for strongly obscured sources.  In the
following we will roughly adopt a completeness limit at
$M=10^{10}M_\odot$ at $z>2$.

\section{The galaxy stellar mass density}

Fig. 1 shows the best--fit stellar mass derived for each galaxy in the
HDFS, at the corresponding redshifts.  We note that, due to the small
volume sampled by the HDFS, the most massive objects have typical
masses around $10^{11}M_\odot$ even at low--intermediate $z$,
nearly a factor of ten smaller than those obtained by larger area
surveys.  We note however that several objects above
$M_*\simeq10^{10}M_\odot$ exist above $z=2$, the most massive being an
ERO at $z_{phot}\simeq 2.8$ with $M_*\simeq 2 \times 10^{11}M_\odot$,
that is better described in a separate paper (Saracco et al.  2003).

To check whether the evolution of the more massive galaxies in our
sample is consistent with theoretical expectations, we have used the
CDM models of Menci et al. 2002 to compute a threshold $M^*_{th}$
defined such that one expects one galaxy above it (in the area sampled
by the HDFS observations) per unit redshift (Fig.1, solid line).  The
curve follows the growth of the high--mass tail of the galaxy stellar
mass function on a hierarchical scenario.
Quantitavely, we find 17 galaxies above the threshold, against the
expected 3, a first evidence that the high-mass tail of the galaxy
stellar mass function is not adequately followed by present CDM
models.

In the upper panel of Fig.2, we present the stellar mass density
$\rho_*(z)$ computed from our whole data sample with the standard
$1/V_{max}$ estimator, corrected to account for incompleteness
analogously to D03.  We plot separately both the SMC and the
Calzetti--based estimates.  We also report in Fig.2 other available
estimates of the stellar mass density as obtained from other surveys
at $z\simeq 1$ (BE00, Cohen 2002), and most
notably with the HDFN results (Dickinson et al. 2003), that has
similar depth, size and adopted technique. While the evolution at
$z\leq 1.5$ in both HDFs is consistent within errors (and with other
surveys), it is remarkable to observe that we find in the HDFS a
stellar mass density at $2\leq z \leq 3.2$ that is about two times
higher than in HDFN.

At face value, the observed values of $\rho_*(z)$ witness a fast
increase of the stellar mass density from 7-15\% of the local (with an
upper limit of 40\%) at $z\geq 2$ to about unity at $z\leq1$, i.e. in
a relatively short amount of cosmic time.  Following Cole et
al. (2000) and D03 , we have compared the observed evolution with
available theoretical expectations.  We use an analytic fit to the
global star--formation rate of Steidel et al. 1999, with two different
dust extinctions ($E(B-V)=0,0.15$), and our CDM hierarchical model:
both the CDM model and the integrated contribution of the global
star--formation rate with a reasonable dust extinction provide a good
fit to the data.

However, a clean interpretation of the HDFS data must take into
account that the sampling of the underlying galaxy stellar mass
function may be incomplete on the massive side, because of the small
HDFS area, and may be inhomogenous at its faint side due to its varying
depth as a function of $z$.  For these reasons, we have also obtained
a homogeneous estimate of its evolution computing the mass density
only at  $10^{10} M_\odot \leq M_* \leq10^{11} M_\odot$
(Fig. 2, lower panel), where our sample is complete and well sampled
at all $z$. This range is slightly below the typical Schecther
mass in the local mass function of Cole et al. 2001. Similarly, we
have also computed the same quantity in the HDF--N field, using the
data of P01 and D03.

On this homogeneous subsample of relatively massive galaxies the
stellar mass density at $z>2$ in the HDFS is $\simeq 20^{+20}_{-5} \%$
of the local value, and approaches the local one at $z\leq 1$. If we
reproduce this selection criteria in our CDM model, that is
overplotted in Fig.2, we find that the present CDM rendition
dramatically fails to reproduce the stellar mass density in massive
galaxies at $z\geq2$. Again, we find that the HDF--N is underabundant
of massive galaxies with respect to HDF--N, but is nevertheless 
well above the CDM predictions at $z\simeq 2.75$.

\section{Star--forming and passive galaxies at $z>2$}

We will analyze here the physical properties of the sample of $K\leq
25$ galaxies at $z_{phot}\geq 2$, that consists of 75 objects,
including the 14 objects with $(J-K)_{AB}>1.34$ emphasized by Franx et
al 2003.

First, we use the $J-K$ color to sample the amplitude of the
rest--frame D4000 break, that is sensitive to the age of the
stellar population  We find that most of the
$K$--selected $z\geq 2$ sample is distributed over a large range  $0
< (J-K)_{AB} \leq 1.5$, suggestive of a significant spread in the stellar
population ages.  This is confirmed by the ages
obtained from the best--fit spectral templates, that are
plotted as a function of $M_*$ in the upper panel of Fig.3.  It is
shown that the observed range of $J-K$ colors translates into a spread
of fitted ages, ranging from $10^{8}$ to $3\times 10^{9}$ Gyrs for the
complete $M_*\gsim 10^{10}M_\odot$ subsample, with about half of the
sample close to the relevant Hubble time. At lower masses, the
fraction of younger objects seems to increase, a result that is likely
affected by the biases against old/passive low mass objects.

At the same time, we find that most of the sample is UV bright (see
Poli et al 2003, Fig.2) and is restricted to the range of $0.1
\lesssim V-I \lesssim 0.5$, corresponding to $0.05 \lesssim E(B-V)
\lesssim 0.2$ for an SMC extinction curve.  We will use in the
following the derived best--fit star formation rates (that include a
correction for reddening), that are in agreement with those derived
with a fixed conversion between the SFR and $L_{1400}$,
assuming an $E(B-V)\simeq 0.15$.


Only seven ``red'' objects are detected at both large $V-I\gsim 1$ and
$J-K\gsim 1.5$, a combination that may be due to either a strongly
absorbed star--forming galaxy or to a passively fading stellar
population, an ambiguity that cannot be safely removed without
spectroscopy (Cimatti et al 2002). For the present dataset, we have
found that the spectral fitting to the complete $UBVIJHK$ distribution
slightly prefers the ``passive'' spectral model: when we force
star--forming dusty solution by selecting $\tau/${\it age}$\geq 2$, we
find that typical $\chi^2_{red}$ are larger by a factor of 2, albeit being
still statistically acceptables ($\chi^2_{red}\leq 1.4$).

The output of the fitting procedure is summarized in the lower panel
of Fig.3, where we plot the specific star--formation rate $\dot M/M_*$
for the $K$--selected $z>2$ sample.  At lower redshifts, the evolution
of $\dot M/M_*$ has been studied by BE00, of which we plot in fig.3
their average relations at $0.2<z<0.5$ and at $0.75<z<1$. They showed
evidence for an increase of the average $\dot M/M_*$ (at a given
$M_*$) with $z$: we find that the trend of increasing $\dot M/M_*$
with $z$ still continues at $z\simeq 3$. A regression to our data
yields $log(\dot M/M_*)=2.18-1.245 log(M_*)$, i.e. the same slope but
a factor 3 higher than the $z\simeq 1$ data. At the median value of
the $M_*> 10^{10} M_\odot$ complete sample ($M_*=3.2\times 10^{10}
M_\odot$) the average specific star--formation rate as derived from
the linear interpolation is $<\dot M/M_*> \simeq 4\times 10^{-10}
$~yr$^{-1}$.  We also note that the adoption of a Calzetti extinction
curve would shift this it to a much higher value of $<\dot M/M_*>
\simeq 2.3\times 10^{-9} $~yr$^{-1}$.  We have found that the average
Scalo $b$ parameter $SFR \times AGE / MASS$, as estimated from the
best--fit quantites, is about 0.8 in this star--forming sample,
suggesting a long duration of the starburst activity.

At the same time, we plot in Fig3 the seven ``red'' objects (over 30
of the mass--complete subsample) discussed before. These objects have
$\dot M/M_* < 10^{-10} $~yr$^{-1}$ i.e. more than 10 times lower that
the typical $z\geq 2$ population if we adopt the
``passively fading'' fitting models, but follow the average trend
if we adopt the ``star--forming'' solution. Given the ambiguity of their
spectral classification, they can be used to obtain an
upper limit to the fraction of galaxies that formed most of their
stars in short episodes prior the time they are observed.  In the
HDF--S, these ``passively fading'' objects contribute at most to about
25\% of the total number density of the mass--complete subsample of
$z\simeq 3$ galaxies. In this case, the cosmologcal mass density due
to these 7 objects is $1.95\times 10^7 M_\odot $~Mpc$^{-3}$, and hence
contributes up to $\simeq 40\%$ of the total mass density of the
mass--complete subsample, that is of $4.5\times 10^7 M_\odot
$~Mpc$^{-3}$. As can be derived from Fig.3, the mass density does not
change appreaciably if we assume the star--forming dusty solution for
these objects.

Finally, we note that most of these ``red'' objects are drawn from the
$(J-K)_{AB}>1.34$ subsample discussed by Franx et al. 2003, although
other objects of this class are actively star--forming. The total mass
density in the $(J-K)_{AB}>1.34$ sample is $3.7\times 10^7 M_\odot
$~Mpc$^{-3}$.

\section{Summary}
We have used new ultradeep $JHK$ images of the HDFS to trace the
evolution of the stellar mass density $\rho_*(z)$ from $z\simeq 0.5$
to $z\simeq3.5$, with the following main results:

- We find clear evidence for a decrease of the observed stellar mass
density with increasing redshifts. At $2\leq z \leq 3.5$ the stellar
mass density of both of the whole $K_{AB}\leq 25$ sample and of the
homogeneous subsample of galaxies with $10^{10}M_\odot \leq M_* \leq
10^{11}M_\odot$ are about 15-20\% (with a solid upper bound of 40\%)
of the local value, a value two times higher than the analogous result
in the HDFN (Dickinson et al. 2003), and approaches the local value
only at $z\simeq 1$.  The UV--based cosmic star--formation history
(e.g. Steidel et al 1999) reproduces the observed evolution of the
total $\rho_*(z)$ (Fig.1).

- In the mass--limited subsample at $z>2$, we find that the fraction of
passively fading galaxies is {\it at most} 25\%, and they can
contribute up to about 40\% of the measured stellar mass density.

- Galaxies of $M_* \simeq 3\times 10^{10}M_\odot$ at $z>2$ form stars
at a specific rate of at least $<\dot M/M_*> \simeq 4\times 10^{-10}
$~yr$^{-1}$, a value 3 times higher than what observed at $z\leq 1$
(Brinchmann and Ellis 2000).  The inverse of this rate is the time
required for these objects to double their mass, assuming constant
star--formation rate: in our sample, this turns out to be 2.5 Gyrs,
comparable to both the Hubble time of the sample and to the cosmic
time up to $z=1$. Given the high fraction of star--forming galaxies in
our mass--complete sample at $z\simeq 2.5$, it is likely that their
observed star--formation episodes last long enough to build up a
substancial fraction of the observed mass density at $z\simeq 1$.

Overall, these data suggest a scenario where the growth of $\rho(z)$
in relatively massive galaxies is consistent with their UV--inferred
star formation properties.

Finally, the large best fit ages and Scalo $b$ parameters of our $z>2$
galaxies suggest that the global star--formation rate should have been
relatively high up to very high $z$, consistent with the results of
recent searches of $z\simeq 5-6$ galaxies (e.g. Fontana et al 2003).

At first glance, this scenario is broadly consistent with the CDM
theoretical expectation, where both ``quiescent'' star--formation rate
and bursts during gas--rich mergings occurr at high rate at $z>1$.
Despite this, our rendition of CDM models largely fails to reproduce
the mass density of the most massive galaxies, even if we assume the
average value between HDF--N and HDF--S.  Hence, the dramatic failure
to reproduce the amount of baryons condensed into stars suggests that
the star--formation in these massive objects occurrs in a more
efficient fashion than what accounted for by our recipes, as also
suggested by other results directly related to the high
star--formation rates found at $z\geq 3$ (Fontana et al. 2003, Poli et
al. 2003).

\newpage

\begin{figure}
\plotone{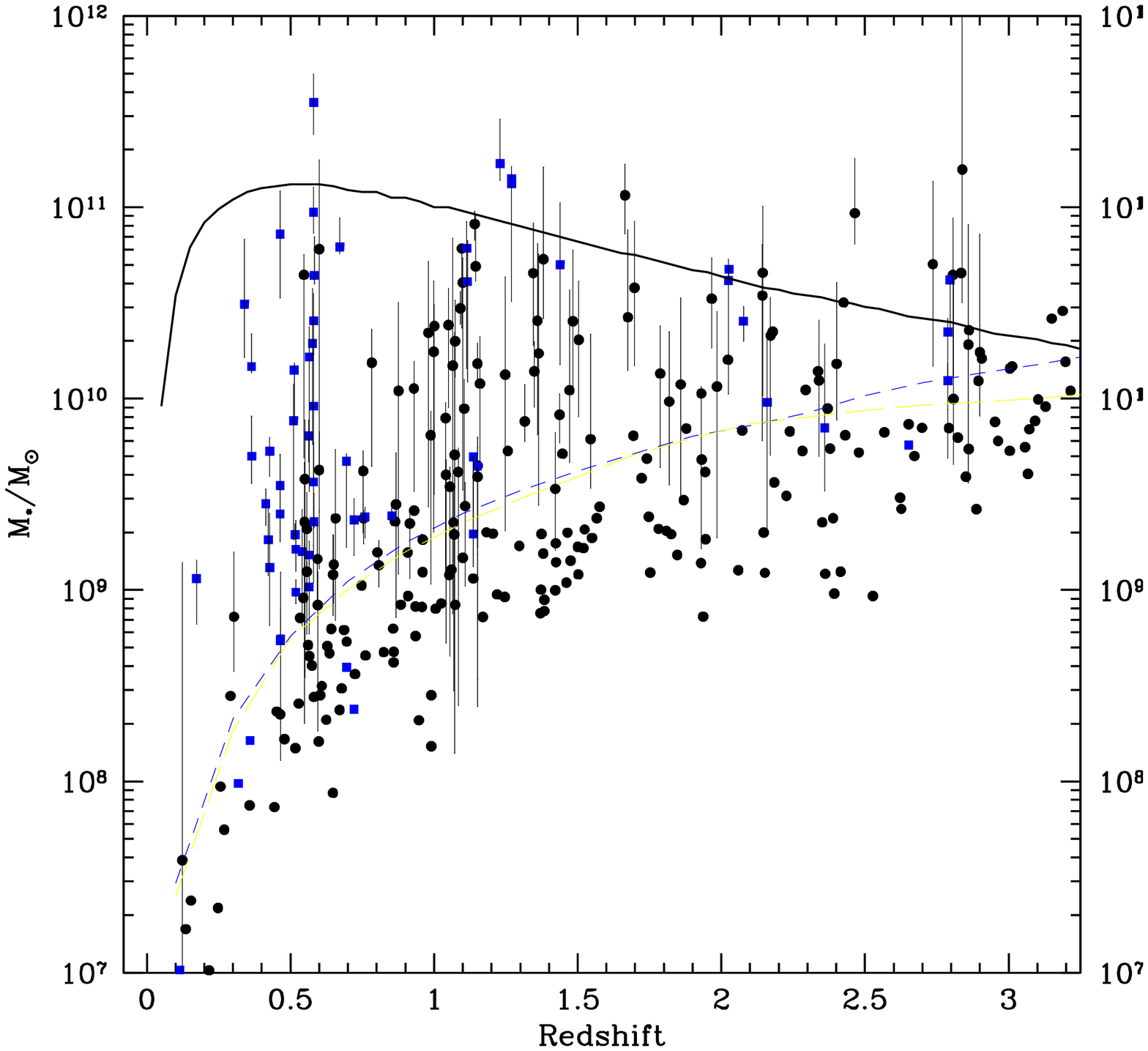}
\label{massz}
\caption { Galaxy stellar masses in the HDFS sample as a function of
z. The points represent the ``best--fit'' estimates with SMC
extinction curve of the stellar masses. Errorbars show the $1-\sigma$
confidence level and are computed including both the SMC and the Calzetti
extinction laws. The solid line has been computed from the CDM
hierarchical models of Menci et al 2002, such that about three objects
are expected from $z=0$ to $z=3$. The short and long dashed line
represent two possible selection curves above which the sample is
mass--complete (see text for details).}
\end{figure}

\newpage

\begin{figure}
\plotone{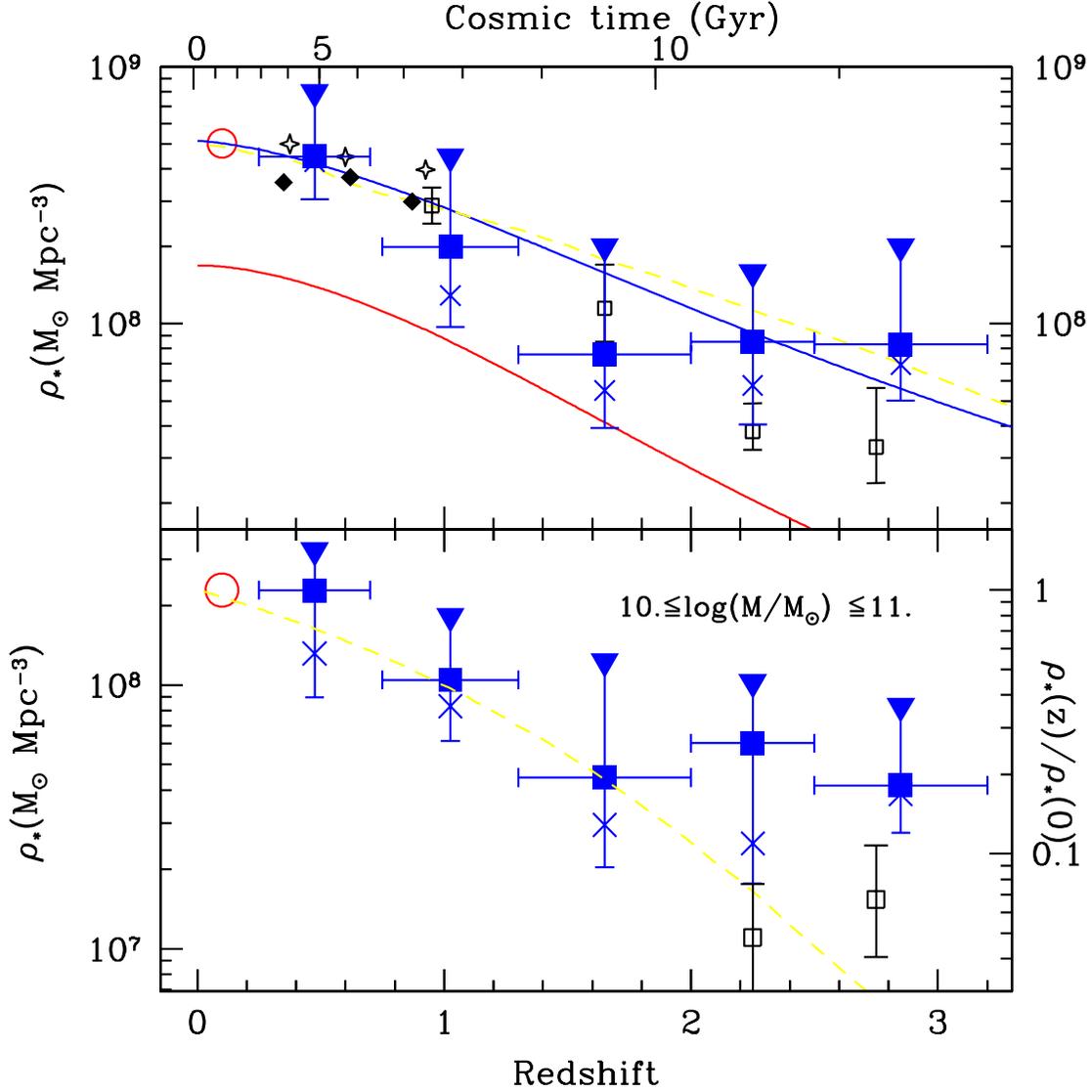}
\label{mdens}
\caption { 
Evolution of the stellar mass density as a function of redshift. {\it
Upper panel:} Observed values as estimated from the present work
(solid squares, SMC; crosses, Calzetti) and from similar surveys:
Dickinson et al 2003 (open squares), Cohen 2002 (open triangles),
Brinchmann and Ellis 2000 (filled triangles), Cole et al 2001
(circle). Upper errorbar are extended at the level obtained from the
``maximal mass'' model, lower represent the number counts noise.  The two
solid lines are the results of integrating the cosmic star formation
history, with and without dust (upper and lower curve,
respectively). The dashed line is the result of the CDM hierarchical
model of Menci et al 2002. {\it Lower panel:} Symbols and curves as in
above, but computed only in the fixed mass interval $10^{10}M_\odot
\leq M_* \leq 10^{11}M_\odot$.}
\end{figure}

\newpage

\begin{figure}
\plotone{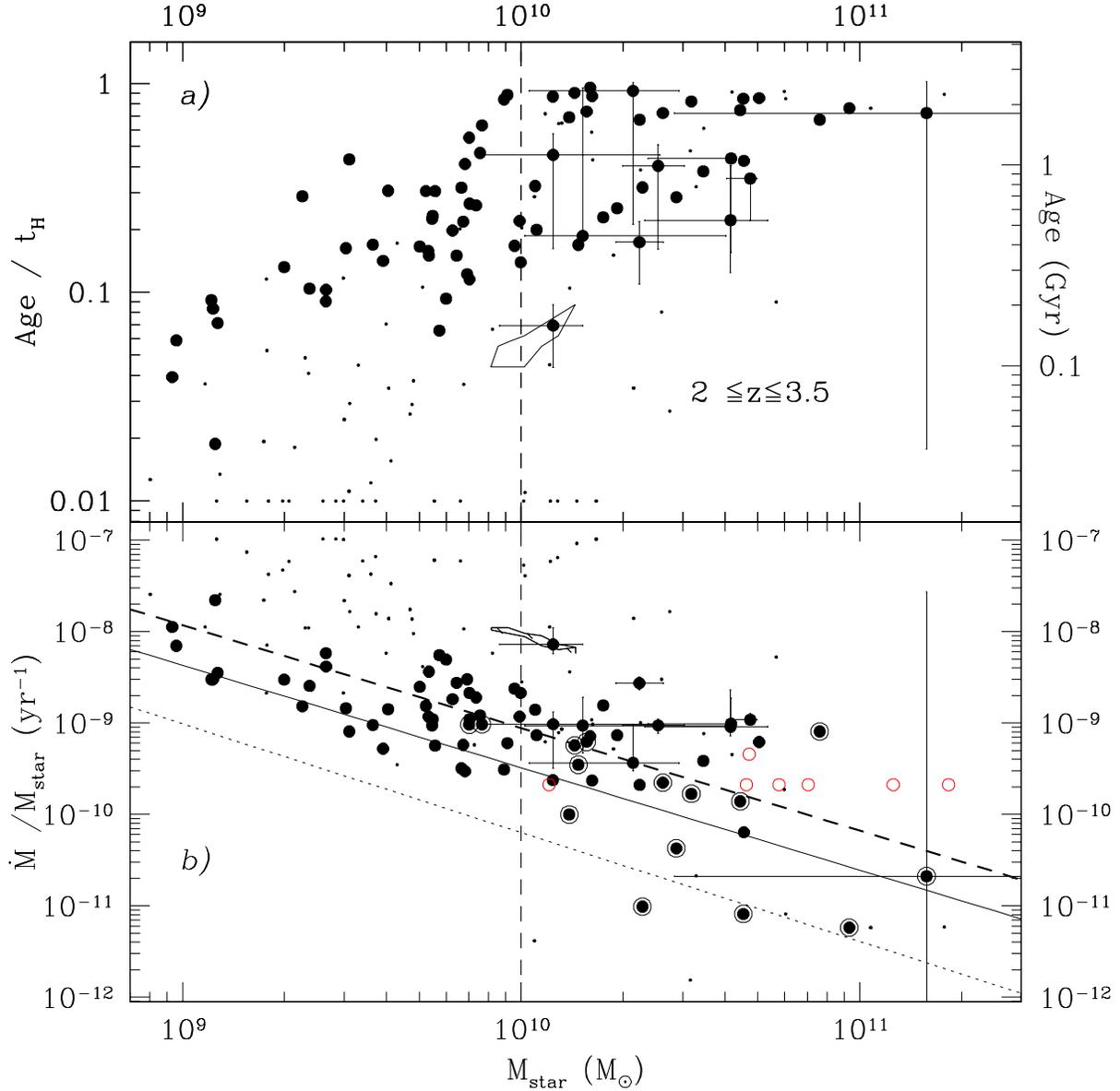}
\label{mod}
\caption{ Rest frame properties of $z>2$ galaxies from spectral
fitting of the HDFS sample. The vertical dashed line show
the rough estimated limit for mass completeness in the $z>2$ sample.
{\it a)} Best--fit age (divided by the corresponding Hubble time) as a
function of stellar mass. For comparison, we show in the right
vertical axis the corresponding ages at $z= 2.5$.  {\it b)}
Specific star--formation rate as a function of stellar mass, compared
with the average relations at $0.2<z<0.5$ (thin dotted line) and at $0.75 <
z < 1$ (solid line) from Brinchmann and Ellis 2000. The thick dashed
line correspond to our fit to the whole SMC sample. In both cases,
filled dots represent the SMC-based estimates, small dots the Calzetti
ones.  Encircled points represent the objects with $J-K\geq 1.34$
discussed by Franx et al 2003.  Errorbars are shown for a few objects,
for the SMC case only: the large errorbars on the most massive objects
reflect the ambiguity in its nature, between passively--fading and
dusty--star forming objects. Note also that errors are typically
correlated, as shown by the $1\sigma$ confidence regions shown for one
of the objects.  Empty points show the estimates of $\dot M/M_*$ when
a star--forming dusty solution is forced for the objects with $\dot
M/M_* \leq 10^{-10} yr^{-1}$: in this case the objects are fitted with
slightly higher masses but without altering the order with respect to mass.}
\end{figure}

\end{document}